\documentclass[aps,prc,superscriptaddress,showpacs,floatfix]{revtex4} 
\usepackage{graphicx,epsfig,amsfonts,amssymb,amsbsy} 
\usepackage[usenames]{pstcol}

\def\openone{\leavevmode\hbox{\small1\kern-3.8pt\normalsize1}}%
 
\def\mf{{\mbox{\tiny\em MFA}}}

\def\bea{\begin{eqnarray}} 
\def\eea{\end{eqnarray}} 
\def\beq{\begin{equation}} 
\def\eeq{\end{equation}}

\begin{document} 
 
\title{Hybrid stars within a covariant, nonlocal chiral quark model} 
\author{D.~B.~Blaschke} 
\email{blaschke@ift.uni.wroc.pl} 
\affiliation{Institute for Theoretical Physics, University of Wroclaw, 
Max Born place 9, 50204 Wroclaw, Poland} 
\affiliation{Bogoliubov  Laboratory of Theoretical Physics, JINR Dubna, 
Joliot-Curie Street 6, 141980  Dubna, Russia} 
\affiliation{Institut f\"ur Physik, Universit\"at Rostock, 
Universit\"atsplatz 3, 18051 Rostock, Germany} 
\author{D.~G\'omez Dumm} 
\email{dumm@fisica.unlp.edu.ar} 
\affiliation{IFLP, CONICET $-$ Dpto.\ de F\'{\i}sica, Universidad Nacional de 
La Plata, C.C. 67, 1900 La Plata, Argentina} 
\affiliation{CONICET, Rivadavia 1917, 1033 Buenos Aires, Argentina} 
\author{A.G. Grunfeld} 
\email{ag.grunfeld@gmail.com} 
\affiliation{CONICET, Rivadavia 1917, 1033 Buenos Aires, Argentina} 
\affiliation{Physics Department, Comisi\'on Nacional de 
Energ\'{\i}a At\'omica, Av.\ Libertador 8250, 1429 Buenos Aires, Argentina} 
\author{T. Kl\"{a}hn} 
\email{thomas.klaehn@googlemail.com} 
\affiliation{Institut f\"ur Physik, Universit\"at Rostock, 
Universit\"atsplatz 3, 18051 Rostock, Germany} 
\affiliation{Gesellschaft f\"ur Schwerionenforschung mbH 
(GSI), 64291 Darmstadt, Germany} 
\affiliation{Physics Division, Argonne National Laboratory,
Argonne, IL 60439-4843, USA} 
\author{N. N. Scoccola} 
\email{scoccola@tandar.cnea.gov.ar} 
\affiliation{CONICET, Rivadavia 1917, 1033 Buenos Aires, Argentina} 
\affiliation{Physics Department, Comisi\'on Nacional de 
Energ\'{\i}a At\'omica, Av.\ Libertador 8250, 1429 Buenos Aires, Argentina} 
\affiliation{Universidad Favaloro, Sol{\'\i}s 453, 1078 Buenos Aires, 
Argentina} 
\date{\today} 
\begin{abstract} 
We present a hybrid equation of state (EoS) for dense matter in which a 
nuclear matter phase is described within the Dirac-Brueckner-Hartree-Fock 
(DBHF) approach and a two-flavor quark matter phase is modelled according 
to a recently developed covariant, nonlocal chiral quark model. We show 
that modern observational constraints for compact star masses ($M\sim 2 
M_\odot$) can be satisfied when a small vector-like four quark interaction 
is taken into account. The corresponding isospin symmetric EoS is 
consistent with flow data analyses of heavy ion collisions and points to a 
deconfinement transition at about 0.55 fm$^{-3}$. 
 
\end{abstract} 
\maketitle 
 
\section{Introduction} 
 
Understanding the properties of matter at moderate and high densities is 
required, e.g., to explain the astrophysical phenomena which accompany the 
birth of compact stars in supernova explosions, and the further evolution 
processes of cooling, spin-down, accretion, merging with companion stars, 
etc., which lead to effects accessible to observation. 
 
Nowadays, one of the questions in the focus of discussions is the 
possibility of a phase transition to deconfined quark matter in the 
stellar cores. There is no doubt that deconfinement of quarks shall occur 
at sufficiently high densities, in accordance with the asymptotic freedom 
of QCD \cite{Collins:1974ky}. 
However, it is not clear a priori whether the critical density for 
deconfinement is low enough to be reached in the cores of neutron stars 
\cite{Baym:1976yu}. 
Even if this condition can be fulfilled, the properties of dense quark 
matter in the vicinity of the deconfinement transition might be too 
similar to those of dense hadronic matter to result in clearly 
distinguishable signals. On the other hand, there have been a number of 
interesting suggestions about how the occurrence of quark matter in 
neutron stars could manifest itself and possibly contribute to the 
resolution of puzzling observations. For the sake of illustration we 
remind on the timing behavior of pulsar spin-down 
\cite{Glendenning:1997fy}, frequency clustering \cite{Glendenning:2000zz} 
or population clustering \cite{Poghosyan:2000mr,Blaschke:2001th} of 
accreting compact stars, which are based on a softening of the equation of 
state and therefore a compactification as well as a reduction of the 
maximum allowable stellar masses. Since deconfined quark matter is rather 
stiff when compared to hyperonic matter, the problem with rather low 
maximum masses of compact stars with hyperonic interior could be solved by 
the occurrence of a quark matter core \cite{Baldo:2003vx}. Similarly, the 
description of the compact star cooling evolution with superconducting 
quark matter interior seems favorable over a purely hadronic 
modelling~\cite{Blaschke:2006gd}. A deconfinement transition during the 
protoneutron star (PNS) evolution offers a mechanism to explain the 
gamma-ray burst energy release of the order of 100~bethe 
(=$10^{53}$~erg)~\cite{Bombaci:2000cv,Berezhiani:2002ks,Aguilera:2002dh}. 
The nucleation timescales for a quark matter phase transition could 
explain the time delay statistics of GRB subpulse structure 
\cite{Drago:2005qb,Drago:2006xa}. In the presence of a strong magnetic 
field, neutrino propagation in hot, superconducting quark matter can 
become collimated (beaming) and asymmetric, thus explaining a resulting 
kick velocity for the PNS \cite{Berdermann:2006rk}. 
 
The recent progress in compact star observations  justifies a reinvestigation 
of the issue of hybrid stars with quark matter cores and theoretical aspects 
of dense matter properties. 
In particular, the high mass of $M=2.1\pm 0.2~M_\odot$ for the pulsar 
J0751+1807 in a neutron star-white dwarf binary system \cite{NiSp05} and  the 
large radius of $R > 12$ km for the isolated neutron star RX J1856.5-3754 
(shorthand: RX J1856) \cite{Trumper:2003we} point to a stiff equation of state 
at high densities. 
Measurements of high masses are also reported for compact stars 
in low-mass X-ray binaries (LMXBs) as, e.g., $M=2.0\pm 0.1~M_\odot$ for 
the compact object in 4U 1636-536 \cite{Barret:2005wd}. 
With data of this kind, new stringent constraints on the equation of state of 
strongly interacting matter at high densities have been formulated (see 
\cite{Klahn:2006ir} and references therein). It has been argued 
\cite{Trumper:2003we,Ozel:2006km} that deconfined quark matter cannot 
exist in the centers of compact stars with masses and radii as reported 
for these objects. In view of recent works on the quark matter EoS, 
however, this claim appears to be premature \cite{Alford:2006vz}. It has 
been demonstrated within the Nambu--Jona-Lasinio model for quark matter 
\cite{Klahn:2006iw} that the inclusion of a diquark condensate (leading to 
color superconductivity) together with a vector meson condensate, not only 
provides a more elaborate description of the EoS but also allows to 
describe the phenomenology of hybrid stars in excellent accordance with 
the above mentioned new mass and mass-radius constraints. The two 
mechanisms at work are: (1) the lowering of the phase transition density 
due to the diquark condensate, so that already typical neutron stars with 
masses in the range $1.1 - 1.5~M_\odot$ can be hybrid stars with extended 
quark matter cores; (2) the stiffening of the EoS due to the vector mean 
field, which implies an increase of the maximum accessible masses of star 
configurations up to $\sim 2~M_\odot$. 
 
Another lesson to be learnt from NJL model studies is that at low 
temperatures there is a sequential deconfinement: strange quarks occur 
only at densities well above the deconfinement of light quarks 
\cite{Gocke:2001ri,Ruster:2005jc,Blaschke:2005uj,Abuki:2005ms,Warringa:2005jh,Warringa:2006dk}.
If those densities could be reached in a compact star, the corresponding 
strange quark matter cores would be in a superconducting CFL phase, which 
renders the hybrid star configuration mechanically unstable 
\cite{Baldo:2002ju,Buballa:2003et,Klahn:2006iw}. 
In the present work we will restrict ourselves to the discussion of the 
two-flavor case, applying the more 
elaborate formalism of a recently developed nonlocal, covariant chiral quark 
model \cite{GomezDumm:2001fz,Duhau:2004pq,GomezDumm:2005hy}. 
Moreover, we consider a 
generalized version of this model, including an isoscalar vector meson 
current which, in the same way as in the case of the Walecka model for 
nuclear matter, leads to a stiffening of the quark matter EoS. 
 
Although we understand hadrons as bound states of quarks, there does not 
yet exist a unified approach which accurately describes the thermodynamics 
of the transition from nuclear to quark matter. Therefore, we apply a 
two-phase description by performing a Maxwell construction describing the 
transition from a nuclear matter EoS to the quark matter 
EoS.  
The nuclear equation of state to be considered here results from 
calculations within the DBHF approach \cite{vanDalen:2004pn} and has 
already been applied for the description of compact stars before 
\cite{Klahn:2006ir,Klahn:2006iw}. Since this EoS is rather stiff, several 
modern compact star observations have been well reproduced. In particular, 
within this approach one is able to obtain large neutron star masses 
($M_{\rm max}=2.33 \,M_\odot$) and radii ($R=12 - 13$ km for typical neutron 
stars). On the other hand, the DBHF description is not so well suited to 
reproduce the results obtained from elliptic flow data in symmetric 
nuclear matter (SNM): from these results, it can be seen that the DBHF EoS 
tends to be too stiff beyond densities of 3 times the saturation density 
$n_{\rm sat}=0.16$ fm$^{-3}$ \cite{Klahn:2006ir}. 
We show here that this problem can be solved by a phase transition to quark 
matter while simultaneously fulfilling 
the constraints on the behavior of dense matter for the case of hybrid stars. 
 
The article is organized as follows: a brief description of the DBHF approach 
and the quark matter model used here is given in Sects.\ II.A and II.B 
respectively. 
In Sect.\ III we show our numerical results, comparing them with present 
empirical constraints on the behavior of dense matter under constraints for 
conditions in neutron stars and heavy ion collisions. 
Our conclusions are stated in Sect.\ IV. Finally, in the 
Appendix we provide some details of the quark matter model.

\section{Theoretical formalism} 
 
As mentioned in the Introduction, in the present calculation we use a two 
phase description to account for the transition from a nuclear matter EoS 
to a quark matter EoS. 
In the following two subsections we briefly {discuss} the theoretical 
approaches considered here {to describe each of these phases.} 
 
\subsection{Nuclear matter equation of state} 
 
For the description of the nuclear matter equation of state we rely on the 
relativistic Dirac-Brueckner-Hartree-Fock (DBHF) approach where the 
nucleon inside the medium is dressed by the self-energy $\Sigma$. 
{This self-energy} is obtained from the Bethe-Salpeter equation for the 
nucleon-nucleon T-matrix in the ladder approximation, using the Bonn-A 
potential in the interaction kernel~\cite{honnef}. 
 
We employ {a} parameterization of the resulting EoS based on 
a parabolic dependence of the energy per nucleon on the asymmetry 
parameter $\alpha = 1 - 2 x$, given in the form 
\begin{equation} 
E(n,\alpha)=E_0(n) + \alpha^2 E_S(n) \ , 
\label{eq:BA} 
\end{equation} 
where  $x=n_p/n$ is the proton fraction, $E_0(n)$ is the energy per 
nucleon in SNM, and $E_S(n)$ is the (a)symmetry energy. 
Both contributions $E_0(n)$ and $E_S(n)$ have been extracted from DBHF 
calculations for the cases $\alpha=0$ and $\alpha=1$, respectively. The 
parabolic interpolation in Eq.~(\ref{eq:BA}) has been widely used in the 
literature, see e.g.\ Refs.~\cite{Klahn:2006ir,Lattimer:2000nx}, and 
proves to be an excellent parameterization of the asymmetry dependence for 
the purpose of the present study. 
An {\em exact reproduction} of  a given 
EoS might require higher order terms, which have been neglected here. 
The advantage of the parabolic interpolation {lies} in the fact that 
all zero temperature equations of state for neutron star matter (NSM) 
can be derived by applying simple thermodynamic identities~\cite{Baldo:1999rq}.
In particular, we obtain 
\begin{eqnarray} 
\varepsilon_B(n,\alpha)&=&nE(n,\alpha) \ , \\ 
P_B(n,\alpha)&=&n^2\frac{\partial}{\partial n} 
E(n,\alpha) \ , \\ 
\label{eq:munp} 
\mu_{n,p}(n,\alpha)&=& 
\left( 
1+n\frac{\partial}{\partial n} 
\right) E_0(n)- 
\left( 
\alpha^2\mp2\alpha-\alpha^2n\frac{\partial}{\partial n} 
\right) E_S(n) \ , 
\end{eqnarray} 
for the baryonic energy density $\varepsilon(n)$, the pressure $P(n)$, and 
the chemical potentials of neutrons $\mu_{n}$ (upper sign) and protons 
$\mu_{p}$ (lower sign), respectively. 
 
\subsection{Quark matter equation of state} 
\label{subsec:QMEoS} 
 
Early reviews on the treatment of quark matter within the NJL model as a chiral
quark model with a local current-current-type interaction can 
be found, e.g., in Refs. \cite{Vogl:1991qt,Klevansky:1992qe,Hatsuda:1994pi}.
In order to include color superconductivity, techniques were needed which are
described, e.g., in Refs. \cite{Berges:1998rc,Buballa:2003qv}.
  
We use here a generalization of these approaches to a nonlocal chiral quark 
model which includes scalar and vector quark-antiquark interactions and 
anti-triplet scalar diquark interactions. 
The corresponding effective Euclidean action in the case of two light flavors 
is given by 
\begin{equation} 
S_E = \int d^4 x \ \left\{ \bar \psi (x) \left(- i \rlap/\partial + m_c 
\right) \psi (x) - \frac{G_S}{2} j^f_S(x) j^f_S(x) - \frac{H}{2} 
\left[j^a_D(x)\right]^\dagger j^a_D(x) {-} 
\frac{G_V}{2} j_V^{\mu}(x)\, j_V^{{\mu}}(x) \right\} \,. 
\label{action} 
\end{equation} 
Here $m_c$ is the current quark mass, which is assumed to be equal for $u$ 
and $d$ quarks, whereas the currents $j_{S,D,V}(x)$ are given by nonlocal 
operators based on a separable approximation to the effective one gluon 
exchange model (OGE) of QCD. These currents read 
\begin{eqnarray} 
j^f_S (x) &=& \int d^4 z \  g(z) \ \bar \psi(x+\frac{z}{2}) \ \Gamma_f\, 
\psi(x-\frac{z}{2})\,, 
\nonumber \\ 
j^a_D (x) &=&  \int d^4 z \ g(z)\ \bar \psi_C(x+\frac{z}{2}) \ i 
\gamma_5 \tau_2 \lambda_a \ \psi(x-\frac{z}{2}) \ ,\nonumber 
\\ 
j^\mu_V (x) &=& \int d^4 z \ g(z)\ \bar \psi(x+\frac{z}{2})\; \gamma_\mu 
\; \psi(x-\frac{z}{2})\ , 
\label{cuOGE} 
\end{eqnarray} 
where we have defined $\psi_C(x) = \gamma_2\gamma_4 \,\bar \psi^T(x)$ and 
$\Gamma_f=(\openone,i\gamma_5\vec\tau)$, while $\vec \tau$ and 
$\lambda_a$, with $a=2,5,7$, stand for Pauli and Gell-Mann matrices acting 
on flavor and color spaces, respectively (notice that $\gamma_\mu = 
(\vec\gamma,\gamma_4)$ are Euclidean Dirac matrices). The functions $g(z)$ 
in Eqs.~(\ref{cuOGE}) are nonlocal covariant form factors characterizing 
the effective quark interaction~\cite{GomezDumm:2005hy}. 
 
The effective action in Eq.~(\ref{action}) might arise via Fierz 
rearrangement from some underlying more fundamental interactions, and is 
understood to be used ---at the mean field level--- in the Hartree 
approximation. In general, the ratios of coupling constants $H/G_S$, 
$G_V/G_S$ would be determined by this microscopic couplings; for example, 
OGE interactions in the vacuum lead to $H/G_S =0.75$ and $G_V/G_S= 0.5$. 
However, since the precise derivation of effective couplings from QCD is 
not known, there is a large theoretical uncertainty in these ratios. 
Details of the values used in the present work will be given below. 
 
We proceed by considering a bosonized version of this quark model, in 
which scalar, vector and diquark fields are introduced. Moreover, we 
expand these fields around their respective mean field values, keeping the 
lowest order contribution to the thermodynamic quantities. The only 
nonvanishing mean field values in the scalar and vector sectors correspond 
to isospin zero fields, $\bar\sigma$ and $\bar\omega$ respectively, while 
in the diquark sector, owing to the color symmetry, one can rotate in 
color space to fix $\bar\Delta_5=\bar\Delta_7=0$, 
$\bar\Delta_2=\bar\Delta$. 
 
Now we consider the Euclidean action at zero temperature and finite baryon 
chemical potential $\mu_B$. Introducing different chemical potentials 
$\mu_{fc}$ for each flavor and color, the corresponding mean field grand 
canonical thermodynamic potential per unit volume can be written as 
\begin{equation} 
\Omega^\mf  =  
\frac{ \bar \sigma^2 }{2 G_S} + \frac{ {\bar \Delta}^2}{2 H} - \frac{\bar 
\omega^2}{2 G_V}\ - \frac{1}{2} \int \frac{d^4 p}{(2\pi)^4} \ \ln 
\mbox{det} \left[ \ S^{-1}(\bar \sigma ,\bar \Delta, \bar \omega, 
\mu_{fc}) \ \right] \ , \label{mfaqmtp} 
\end{equation} 
where the inverse propagator $S^{-1}$ is a $48 \times 48$ matrix in Dirac, 
flavor, color and Nambu-Gorkov spaces \cite{GomezDumm:2001fz,Duhau:2004pq}. 
Its explicit regularized form $\Omega_{\rm (reg)}^\mf$ is given in the 
Appendix, together with further details about the model. The mean field 
values $\bar \sigma$, $\bar \Delta$ and $\bar \omega$ are obtained from 
the coupled equations 
\begin{eqnarray} 
\frac{ d \Omega^\mf}{d\bar \Delta} \ = \ 0 \ , \ \ \ 
\frac{ d \Omega^\mf}{d\bar \sigma} \ = \ 0 \ , \ \ \ 
\frac{ d \Omega^\mf}{d\bar \omega} \ = \ 0 \ . 
\label{gapeq} 
\end{eqnarray} 
 
In principle one has six different quark chemical potentials, 
corresponding to quark flavors $u$ and $d$ and quark colors $r,g$ and $b$. 
However, there is a residual color symmetry (say, between red and green 
colors) arising from the direction of $\bar\Delta$ in color space. 
Moreover, if we require the system to be in chemical equilibrium, it can 
be seen that chemical potentials are not independent from each other. 
In general, it is shown that all $\mu_{fc}$ can be written in terms of three 
independent quantities: the baryonic chemical potential $\mu_B$, a quark 
electric chemical potential $\mu_{Q_q}$ and a color chemical potential 
$\mu_8$. 
The corresponding relations read 
\begin{eqnarray} 
\mu_{ur} = \mu_{ug} &=& \frac{\mu_B}{3} + \frac23 \mu_{Q_q} + \frac13 
\mu_8 \ \ ,  \qquad \mu_{ub} = \frac{\mu_B}{3} + \frac23 \mu_{Q_q} - 
\frac23 \mu_8 \ \ , 
\nonumber \\ 
\mu_{dr} = \mu_{dg} &=& \frac{\mu_B}{3} - \frac13 \mu_{Q_q} + \frac13 
\mu_8 \ \ ,  \qquad \mu_{db} = \frac{\mu_B}{3} - \frac13 \mu_{Q_q} - 
\frac23 \mu_8 \ \ . 
\label{chemical} 
\end{eqnarray} 
The chemical potential $\mu_{Q_q}$, which distinguishes between up 
and down quarks, as well as the color chemical potential $\mu_8$, which 
has to be introduced to ensure color neutrality 
\cite{Iida:2000ha,Alford:2002kj,Steiner:2002gx}, vanish for an isospin 
symmetric quark matter system. 
Thus, in this case, the corresponding EoS can be obtained after calculating 
the mean field values $\bar\sigma$, $\bar\omega$ and $\bar\Delta$ from 
Eqs.~(\ref{gapeq}). 
 
Now, if we want to describe the behavior of quark matter in the core of 
neutron stars, in addition to quark matter we have to take into account 
the presence of electrons and muons. 
Thus, treating leptons as a free relativistic Fermi gas, the total pressure 
of the quark matter + lepton system is given by 
\begin{equation} 
P = - \;\Omega^\mf_{\rm (reg)}\; -\; \Omega^l \ , 
\label{potential} 
\end{equation} 
where $\Omega^l$ is the thermodynamical potential per unit volume for a gas 
of noninteracting electrons and muons (see Appendix A). 
In addition, it is necessary to take into account that quark matter has to be 
in $\beta$ equilibrium with electrons and muons through the $\beta$ decay 
reactions 
\begin{equation} 
d\to u+l+\bar\nu_l\ , \qquad u+l\to d+\nu_l\ , 
\end{equation} 
for $l=e,\mu$. 
Thus, assuming that (anti)neutrinos escape from the stellar core, we have 
an additional relation between fermion chemical potentials, namely 
\begin{equation} 
\label{betaeq} 
\mu_{dc} - \mu_{uc} = - \mu_{Q_q} = \mu_l 
\end{equation} 
for $c=r,g,b$, $\mu_e = \mu_\mu = \mu_l$. 
 
Finally, in the core of neutron stars we also require the system to be 
electric and color charge neutral, hence the number of independent 
chemical potentials reduces further. 
Indeed, $\mu_l$ and $\mu_8$ get fixed by the condition that charge and 
color densities vanish, 
\begin{eqnarray} 
\rho_{Q_{tot}} &=& \rho_{Q_q}- \sum_{l=e,\mu}\rho_l 
\ = \ \sum_{c=r,g,b} \left(\frac23 \ \rho_{uc} - \frac13 \ \rho_{dc} \right) 
- \sum_{l=e,\mu}\rho_l \ = \ 0 \ \ , \nonumber \\ 
\rho_8 & = & \frac{1}{\sqrt3} \sum_{f=u,d} 
\left(\rho_{fr}+\rho_{fg}-2\rho_{fb} \right) \ = \ 0 \ , 
\label{dens} 
\end{eqnarray} 
where the expressions for the lepton densities $\rho_l$ and the quark 
densities $\rho_{fc}$ can be found in the Appendix. 
Note that the set of color chemical potentials required to ensure color
neutrality of the system depends on the choice of the orientation of the 
diquark condensate orientation in color space \cite{Buballa:2005bv}.
For the standard choice employed in the present work, $\mu_8$ is sufficient.
In summary, in the case of neutron star quark matter, for each value of 
$\mu_B$ one can find the values of $\bar \Delta$, $\bar \sigma$, $\omega$, 
$\mu_l$ and $\mu_8$ by solving Eqs.~(\ref{gapeq}), supplemented by 
Eqs.~(\ref{betaeq}) and (\ref{dens}). 
This allows to obtain the quark matter EoS in the thermodynamic region we 
are interested in. 
 
\section{Numerical results and discussion} 
 
In this section we present our numerical results, showing the behavior of 
both isospin symmetric hadronic matter and neutral hadronic matter for 
finite baryochemical potential. 
The nuclear-to-quark matter phase transition is treated in the traditional 
way, following a two-phase scheme in which the nuclear and quark matter phases 
are described by the theoretical approaches presented in the previous section. 
 
{As stated, the nuclear matter phase is described according to the DBHF 
approach, in which the nucleon self-energy is calculated from the 
Bethe-\-Salpeter equation considering a Bonn-A potential in the interaction 
kernel.} 
Regarding the quark matter model, we note first that, according to previous 
analyses carried out within nonlocal scenarios
\cite{GomezDumm:2001fz,Duhau:2004pq}, the results are not expected to show a 
strong qualitative dependence on the shape of the nonlocal form factors. 
Thus we will consider (in momentum space) a simple and well-behaved Gaussian 
function, 
\begin{equation} 
g(p^2)  =  \exp(-p^2/\Lambda^2) \ , 
\label{reg2} 
\end{equation} 
where $\Lambda$ is a free parameter of the model, playing the role of an 
ultraviolet cut-off. 
The value of $\Lambda$, as well as the values of the free model parameters 
$G_S$ and $m_c$, can be fixed from low energy phenomenology. 
Here we have chosen these input parameters so as to reproduce the empirical 
values for the pion mass $m_\pi=139$ MeV and decay constant $f_\pi=92.4$ MeV, 
and to obtain a phenomenologically reasonable value for the chiral condensate 
at vanishing $\mu_B$, namely $\langle 0|\bar q q|0\rangle^{1/3} = - 230$~MeV. 
In this way we obtain $m_c = 6.49$~MeV, $G_S = 0.515\times 10^{-4}$~MeV$^{-2}$ 
and $\Lambda = 678$~MeV~\cite{GomezDumm:2006vz}. 
 
The values of the scalar diquark coupling and the isoscalar vector coupling 
$H$ and $G_V$, or equivalently the dimensionless ratios $h=H/G_S$ and 
$g=G_V/G_S$, are considered here as parameters to be chosen in 
accordance with phenomenological constraints from flow data analyses of 
heavy-ion collisions and the new mass and mass-radius constraints from 
compact star observations. 
In accordance with our previous investigation of the phase diagram of neutral 
quark matter in Ref.~\cite{GomezDumm:2005hy} we choose two typical values 
$h=0.70$ and $h=0.74$ for the diquark/scalar coupling ratio, and analyze the 
phase transition features for different values of $g$ so as to obtain an 
acceptable transition density. 
 
\begin{figure} 
\begin{tabular}{cc} 
\includegraphics[width=0.42\textwidth,angle=-90,clip=]{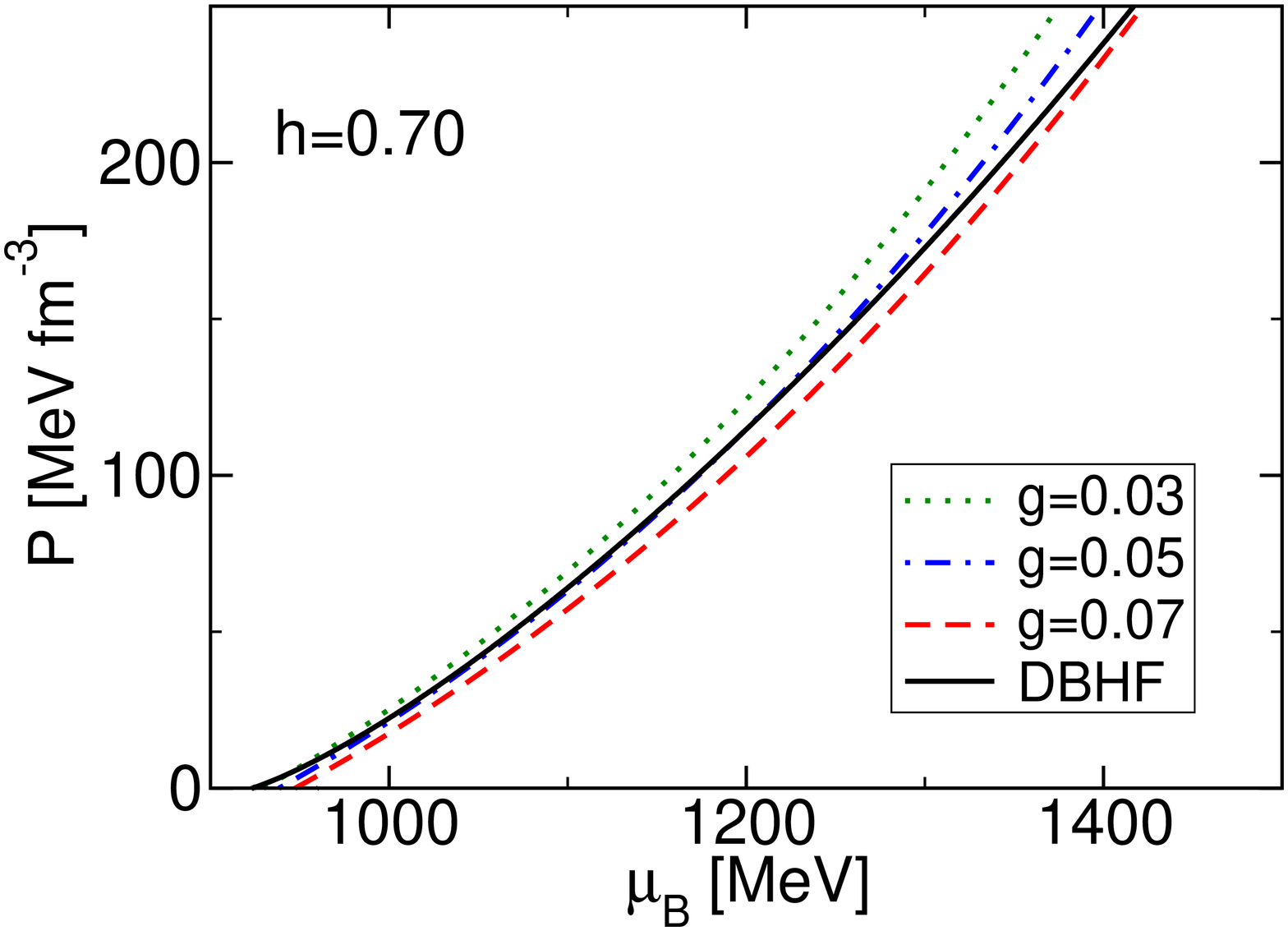}& 
\hspace{-10mm} 
\includegraphics[width=0.42\textwidth,angle=-90,clip=]{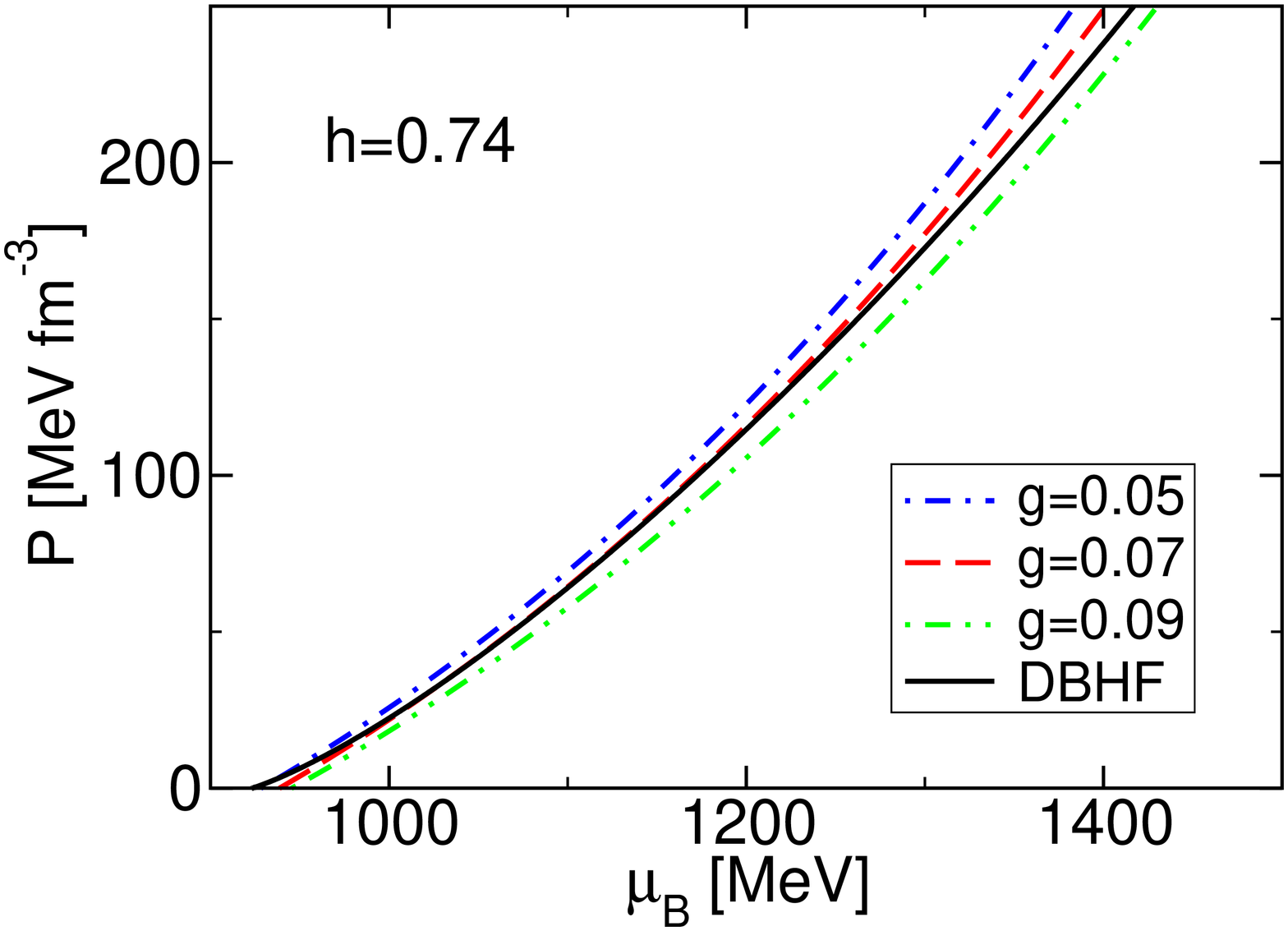} 
\end{tabular} 
\caption{(Color online) Pressure as a function of the baryochemical 
potential for isospin symmetric matter. The nuclear matter phase is 
modelled by the DBHF equation of state with the Bonn-A potential 
\cite{vanDalen:2004pn} (solid line) and the results for the covariant, 
nonlocal chiral quark model are given for different scaled vector coupling 
strengths $g=0.03,~0.05,~0.07,~0.09$ (dotted, dash-dotted, dashed and 
dash-double-dotted curves, respectively) and scaled diquark coupling 
strengths of $h=0.70$ (left panel) and $h=0.74$ (right panel). A phase 
transition to quark matter is obtained at the crossing of nuclear and 
quark matter curves, for discussion see text.} \label{P-mu} 
\end{figure}

\begin{figure} 
\includegraphics[width=0.45\textwidth,angle=-90]{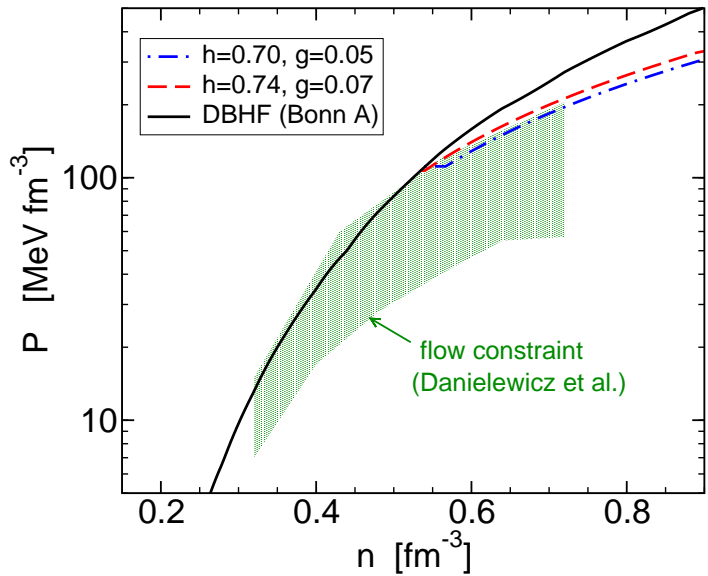} 
\caption{(Color online) Pressure as a function of the density for 
isospin symmetric matter. The phase transition to quark matter 
softens the EoS at densities above $0.55$ fm$^{-3}$, relative to the 
nuclear  DBHF EoS, thus allowing to fulfill the flow constraint derived in 
Ref.~\cite{Danielewicz:2002pu}.  Line styles as in Fig.~\ref{P-mu}.} 
\label{P-n} 
\end{figure}

In Fig.~\ref{P-mu} we show the curves for the pressure as function of the 
baryochemical potential, for both nuclear and quark matter phases, in the 
case of isospin symmetric matter. Left and right panels correspond to 
$h=0.70$ and $h=0.74$ respectively, and in each case three values of the 
scaled vector coupling $g$ have been chosen. As can be seen from the 
figures, within these ranges of $g$ both nuclear and quark matter EoS 
behave similarly in the relevant domain of baryochemical potentials, 
therefore the value of the critical $\mu_B$ strongly depends on the value 
of the vector coupling. In any case, for both $h=0.70$ and $h=0.74$ it is 
possible to tune the value of $g$ so that the softening due to the 
deconfinement transition occurs at a density of about $0.55$~fm$^{-3}$. 
At this point the DBHF EoS becomes too stiff, being unable to fulfill the 
flow constraint~\cite{Danielewicz:2002pu}. This situation is sketched in 
Fig.~\ref{P-n}, where we show the area allowed by the flow constraint and 
the corresponding DBHF curve in the density-pressure plane. For $h=0.70$ 
and $h=0.74$, the values $g=0.05$ and $g=0.07$ respectively (dotted lines 
in the $P-\mu_B$ curves of Fig.~\ref{P-mu}) lead to nuclear-to-quark 
matter phase transitions such that the stiffness of the pressure curve is 
softened and the flow constraint can be satisfied. On the other hand, 
values of $g$ outside the range considered in Fig.~\ref{P-mu} would lead 
to either a too early phase transition or to a situation in which there is 
no transition to quark matter at all. We emphasize that both the nuclear 
and quark matter EsoS behave almost identically over a wide range of the 
baryochemical potential in Fig.~\ref{P-mu}. 
This results in an almost direct {\em crossover} transition with a small 
or even negligible density jump at the transition density, as shown in 
Fig.~\ref{P-n}. 
As a further consequence, the transition density strongly depends on small 
changes of $g$. 
 
\begin{figure} 
\includegraphics[width=0.45\textwidth,angle=-90]{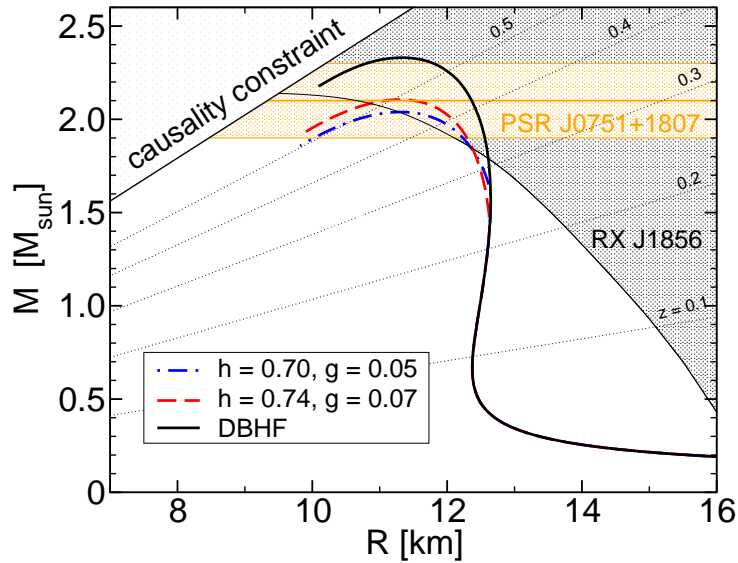} 
\caption{(Color online) Mass-radius relationships for neutron star 
configurations (DBHF EoS, solid line) and hybrid star 
configurations with hadronic shell (DBHF EoS) and color superconducting 
quark matter core (nonlocal chiral quark model EoS) for {two} 
parameter sets characterized by the coupling ratios $h=H/G_S$ and 
$g=G_V/G_S$. Dashed line: $h=0.74$, $g=0.07$; dash-dotted line: $h=0.70$, 
$g=0.05$.} 
\label{M-R-Z} 
\end{figure}

Finally, let us take into account the hadronic matter models leading to 
the curves in Fig.~\ref{P-n}, considering now the description of {\em 
neutral} hadronic matter in the interior of compact stars. In 
Fig.~\ref{M-R-Z}, we show the sequences of compact star configurations 
obtained as solutions of the Tolman-\-Oppenheimer-\-Volkoff equations of 
general-relativistic hydrodynamic stability, for selfgravitating dense 
hadronic matter described by these theoretical models. The results quoted 
in Fig.~\ref{M-R-Z} represent the main outcome of this work: using the 
quark matter EoS derived from a covariant nonlocal chiral quark model, 
generalized here by including a vector interaction, we obtain hybrid star 
configurations which fulfill the modern constraints on high masses and 
radii of compact stars discussed in the Introduction. Moreover, the 
deconfinement transition for isospin symmetric matter occurs at about 
threefold nuclear saturation density, and results in a sufficient 
softening of the EoS thus circumventing a violation of the flow constraint 
from heavy-ion collisions. 
 
\section{Conclusions} 
 
We have generalized in this work a recently developed covariant, nonlocal 
chiral quark model by including a vector-vector four quark interaction 
which leads to a stiffening of the corresponding quark matter EoS. This 
allows us to describe compact stars possessing a quark matter core {\em 
and} being in agreement with the modern compact star phenomenology, which 
suggests stars with maximum masses of $\sim 2~M_\odot$ and radii of $12 - 
13$ km. 
 
We show that the traditional application of Gibbs conditions for phase 
equilibrium to construct a phase transition between deconfined quark 
matter {--} described in the new approach developed here {--} 
and nuclear matter {-- described} by the DBHF approach {--} with 
the Bonn-A potential, is very sensitive to tiny changes of parameter 
values: quark and hadronic matter EsoS behave rather similarly in the 
vicinity of the phase transition and lead to a crossover-like behavior for 
the hybrid EoS. This reconfirms an earlier discussed 'masquerade' effect 
\cite{Alford:2004pf,Klahn:2006iw,Blaschke:2006gd} for hybrid stars within 
the present approach. After exploring the allowed range of model 
parameters, we find values for which both the flow constraint from 
heavy-ion collisions and the mass constraint for hybrid stars are 
satisfied. 
 
We want to point out that the deconfinement phase transition should be a 
rather robust phenomenon of the high-density EoS. Therefore, the mentioned 
'masquerade' effect may point to a deficiency in the two phase approach 
for the description of hybrid EsoS. It is a demanding task to develop 
unified approaches for quark/nuclear matter on the basis of chiral quark 
models in which nucleons and mesons appear as relativistic bound states of 
quarks and antiquarks (for first steps in this direction, see 
Ref.~\cite{Lawley:2006ps,Rezaeian:2006yj}). Under conditions of high 
density and/or temperature these bound states dissociate into continuum 
correlations (resonances) in quark matter within a Mott transition (see 
\cite{Ropke:1986qs} for a model calculation within a nonrelativistic Green 
functions approach, which has also lead to an early prediction of stable 
quark matter cores in compact stars~\cite{Blaschke:1989nn}). 
 
We believe that the covariant nonlocal chiral quark model presented here 
could be developed to a more elaborate approach unifying quark and 
hadronic matter descriptions on the quark level. The realization of such a 
project is beyond the scope of the present work. 
 
\section*{Acknowledgments} 
 
This work has been supported in part by CONICET and ANPCyT (Argentina), 
under grants PIP 6009, PIP 6084 and PICT04-03-25374, and by a scientist 
exchange program between Germany and Argentina funded jointly by DAAD and 
ANTORCHAS under grants No. DE/04/27956 and 4248-6, respectively. 
T.K. acknowledges support 
by the GSI Darmstadt and D.B.B. by the Polish Committee of Scientific Research.
 
\begin{appendix} 
\section{Details of the nonlocal model for quark matter} 
 
In this Appendix we show some explicit expressions corresponding to the 
nonlocal chiral quark model considered in Subsect. \ref{subsec:QMEoS}. The 
determinant of the $48\times 48$ matrix $S^{-1}$ appearing in 
Eq.~(\ref{mfaqmtp}) can be analytically calculated. In this way one 
obtains 
\begin{eqnarray} 
\Omega^\mf &=& \frac{\bar \sigma^2}{2 G_S}\ +\ \frac{\bar \Delta^2}{2 H}\ 
- \frac{\bar \omega^2}{2 G_V}\ -\ \int \frac{d^4 p}{(2\pi)^4} 
\sum_{c\,=\,r,g,b} \ln |A_c|^2 \ , \label{once} 
\end{eqnarray} 
where 
\begin{equation} 
A_c \ = \ \Big[({\tilde p}_{uc}^{\,+})^2 + 
({\Sigma^p_{uc}})^2\Big]\; \Big[ ( {\tilde p}_{dc}^{\,-})^2 + 
\left({\Sigma^p_{dc}}^\ast\right)^2\Big] \; + \;(1-\delta_{bc})\; 
{\Delta^p}\,^2\; \Big[{\Delta^p}\,^2 + 2\, \tilde p_{uc}^{\,+} 
\cdot {\tilde p}_{dc}^{\,-} + 2\, \Sigma^p_{uc} \, 
{\Sigma^p_{dc}}^\ast\Big] \ , \label{def1} 
\end{equation} 
with the following definitions: 
\begin{eqnarray} 
{\tilde{p}_{fc}}^{\;\pm} & = & \left(\vec p\ , \; p_4 \mp i 
\,[\,\mu_{fc} - 
\bar \omega \, g(p^{\,\pm\;2}_{fc})\,] \right) \ , \nonumber \\ 
p_{fc}^{\,\pm} & = & \left( \vec p\, , \; p_4 \, \mp \, i \,\mu_{fc} \right) \ , \nonumber \\ 
\Sigma^p_{fc} & = & m_f \; + \; \bar \sigma \; g(p^{+\;2}_{fc})\ , \\ 
\Delta^p & = & \bar \Delta \ g\bigg(\frac{[p^+_{ur} + 
p^-_{dr}]^2}{4}\bigg) \ , 
\end{eqnarray} 
where $f = u,d$, and $c=r,g,b$. Notice that due to the symmetry between 
red and green colors one has $\mu_{fr} =\mu_{fg}$. 
 
In general, for finite values of the current quark mass, $\Omega^\mf$ 
turns out to be divergent. We have used here a regularization procedure in 
which we add and subtract the thermodynamical potential for a free quark 
gas, namely 
\begin{equation} 
\Omega^\mf_{\rm (reg)} \  =  \ \Omega^\mf \ - \ 
\Omega^{\rm free}_{\rm (nonreg)} \ + \ \sum_{f,c}\Omega^{\rm 
free}_{{\rm (reg)}fc}\ , \label{omegareg} 
\end{equation} 
In the right hand side, $\Omega^{\rm free}_{\rm (nonreg)}$ is obtained 
from Eq.~(\ref{once}) just by setting $\bar \Delta = \bar \sigma = \bar 
\omega=0$, while for each fermion species $i$ the regularized free 
thermodynamical potential $\Omega^{\rm free}_{{\rm (reg)}i}$ is given by 
\begin{equation} 
\Omega^{\rm free}_{{\rm (reg)}i} \ = \ -\;\frac{1}{24\pi^2}\ 
m_i^4 \ F(\mu_i/m_i) \ , 
\label{freepot} 
\end{equation} 
with 
\begin{equation} 
F(x) \ = \ 2\,x\,(x^2-5/2)\,\sqrt{x^2-1} 
\;+\; 3\,\ln \big(x+\sqrt{x^2-1}\,\big) \ . 
\label{freeomegareg} 
\end{equation} 
The sum in Eq.~(\ref{omegareg}) extends over all quark flavors and colors. 
In the case of leptons, since they can be treated as free particles, the 
corresponding contribution to the compact star thermodynamical potential 
is simply given by 
\begin{equation} 
\Omega^l \ = \ \sum_l \Omega^{\rm free}_{{\rm (reg)}l} \ , 
\end{equation} 
with $l=e,\mu$. 
 
Finally, according to our regularization prescription, the fermion densities 
appearing in Eq.~(\ref{dens}) are given by 
\begin{eqnarray} 
\rho_{fc} & = & -\; \frac{ \partial \Omega^\mf}{\partial \mu_{fc}} \; 
-\; \frac{ \partial \Omega^{\rm free}_{\rm (nonreg)}}{\partial \mu_{fc}} 
\; + \; {\rho_{\rm (reg)}^{\rm free}}_{fc} \ , \nonumber \\ 
\rho_l & = & {\rho_{\rm (reg)}^{\rm free}}_l \ , 
\label{densities} 
\end{eqnarray} 
where $f=u,d$, $c=r,g,b$, and $l=e,\mu$. The fermion density ${\rho_{\rm 
(reg)}^{\rm free}}_i$ of a free particle gas can be easily obtained from 
the regularized thermodynamical potential in Eq.~(\ref{freepot}), yielding 
\begin{eqnarray} 
{\rho_{\rm (reg)}^{\rm free}}_i \ = \ \frac{1}{3\pi^2} \;\, 
(\mu_i^2 - m_i^2)^{3/2} \; \ , 
\end{eqnarray} 
while explicit expressions for the partial derivatives in 
Eq.~(\ref{densities}) can be obtained from the results quoted in 
Ref.~\cite{GomezDumm:2005hy}, taking the $T=0$ limit. 
 
\end{appendix}

\end{document}